\documentclass[10pt, conference, compsocconf]{IEEEtran}
\usepackage{amsmath, amssymb}
\usepackage{algorithm}
\usepackage{algorithmic}
\usepackage{graphicx}

\begin{document}
\title{Delay Modelling for Single Cell IEEE 802.11 WLANs Using a Random Polling System}

\author{
\IEEEauthorblockN{Albert Sunny\IEEEauthorrefmark{1}, Joy Kuri\IEEEauthorrefmark{2} and Saurabh Aggarwal\IEEEauthorrefmark{3}}
\IEEEauthorblockA{Center for Electronics Design and Technology\\
Indian Institute of Science, Bangalore-560012, India\\
Email: \IEEEauthorrefmark{1}salbert@cedt.iisc.ernet.in, \IEEEauthorrefmark{2}kuri@cedt.iisc.ernet.in, \IEEEauthorrefmark{3}saggarwal@cedt.iisc.ernet.in}
}

\bibliographystyle{IEEEtran}

\maketitle

\begin{abstract}
In this paper, we consider the problem of modelling the average delay experienced by a packet in a single cell IEEE 802.11 DCF wireless local area network. The packet arrival process at each node $i$ is assumed to be Poisson with rate parameter $\lambda_i$. Since the nodes are sharing a single channel, they have to contend with one another for a successful transmission. The mean delay for a packet has been approximated by modelling the system as a 1-limited Random Polling system with zero switchover time. We show that even for non-homogeneous packet arrival processes, the mean delay of packets across the queues are same and depends on the system utilization factor and the aggregate throughput of the \textit{MAC}. Extensive simulations are conducted to verify the analytical results.	
\end{abstract}


\IEEEpeerreviewmaketitle

\section{\large{Introduction and Related Work}}
\label{intro}

The IEEE 802.11 has become ubiquitous and gained widespread popularity as a protocol for wireless networks. As a result, various models have been proposed to analyze and model the parameters of interest.

Since the seminal paper by Bianchi \cite{bianchi}, throughput analysis of IEEE 802.11 DCF has come under much scrutiny. In \cite{bianchi}, the author evaluates the aggregate system throughput as a function of the number of nodes under saturation, i.e., when each user has a packet to transmit at all times. The main feature of the analysis is the 2-dimensional Markov model, which captures the back-off phenomenon of IEEE 802.11, given a transmission attempt rate for each node. Due to the robustness and simplicity of the model, it has been used extensively by various researchers.  In \cite{tay}, the authors give an analytical model for throughput analysis of DCF using average back-off state as compared to the Markovian model being proposed by Bianchi. Although the approaches are different, the end numerical results are close to each other. In \cite{akumar}, the authors study the fixed point solution and performance measure in a more generalized framework.

Delay analysis of IEEE 802.11 DCF is limited in comparison to the throughput studies. In \cite{mmc}, the authors present delay analysis of an HOL packet for the saturated scenario. In \cite{tickoo} and \cite{sikdar}, the authors extended the model of Tay and Chua \cite{tay} by proposing G/G/1 queues for each individual user. However, the analysis ignores the random delays due to packet transmissions by other users. The authors arrive at an expression for unsaturated collision probability using fixed point analysis and use the same in their subsequent modelling. 

In \cite{tobagi}, the authors propose System Centric and User Centric Queuing Models for IEEE 802.11 based Wireless LANs. \cite{tobagi} assumes the server to allocate its resources to users in a round robin manner. In the System Centric Model, the arrivals are assumed to be Poisson, thus the resource sharing model takes the form of an M/G/1/PS system with the mean delay being the same as that in an equivalent M/M/1 system. In the User Centric  Model, each user queue is modeled as a separate G/G/1 queue. 

A novel model based on diffusion approximations has been used to model delay in  ad-hoc networks  by Bisnik and Abouzeid \cite{bisnik}. The authors provide scaling laws for delay under probabilistic routing, routing is oblivious to the origin and nature of the packet. In the paper, the authors consider the problem of characterizing average delay over various network deployments. But for a given network deployment, the value of the observed average delay may vary widely in comparison to the value as calculated using the diffusion approximation model. We are interested in a simple model to obtain average delay for a given mesh network as against the average over many random deployments \cite{bisnik}.

In \cite{panda}, the authors provide an analysis of the coupled queue process by studying a lower dimensional process and by introducing a certain conditional independence approximation. But the authors provide an analytical framework to model the delay for the case of homogeneous Poisson arrivals only. In \cite{joy}, we have analyzed the mean delay for single hop wireless mesh networks under light aggregate traffic. In \cite{joy}, assuming constant throughput, we model the system as decoupled queues which receive the same share of the aggregate throughput. We derive a simple closed form expression for the upper bound of delay under homogeneous Poisson packet arrival. We have also describe briefly, the approach to model delay in case of non-homogeneous Poisson process under light to moderate load regime. Through simulations, we show that the computed mean delay and simulated values are close to each other under light aggregate load. But as load increases, interactions between the queues appear and our modelling assumption ceases to be valid. 

In this paper, we model the system as a 1-limited random polling system with zero switchover time. This enables us to use the mean delay expressions from \cite{lee} to analyze the delay in a single cell wireless local area network. We remark that the user traffic delay is not merely the Head-Of-Line (HOL) packet delay that has been analyzed in \cite{bianchi}, \cite{tay} and \cite{akumar}; it includes the delay from the time a user packet arrives at the queue, till the packet leaves the node. Thus, both queuing delay and HOL delay are included. 

Our objective is to explore the use of known results for the \emph{saturated} network and \emph{random polling systems} in analyzing the mean delay experienced by a packet. 
\begin{itemize}
\item We propose a random polling system framework to analyze mean delay in a single cell wireless local area network.
\item We obtain closed form expressions for mean delay by applying results from \cite{lee} to our random polling system framework.
\item We show though simulations that our random polling framework can be used to estimate the mean delay in a single cell IEEE 802.11 wireless local area network in the entire capacity region.
\end{itemize}

The rest of the paper is organized as follows. In Section II, the system model is described in detail. In Section III, we present the random polling system framework (RPS) and obtain a closed form expression for mean delay under Poisson packet arrivals. In Section IV, the proposed framework is validated against simulation results. Finally, Section V presents conclusion and remarks regarding future work.

\section{\large{System model}}
\label{model}
\begin{figure}[h]
\centering
\includegraphics[scale=0.4]{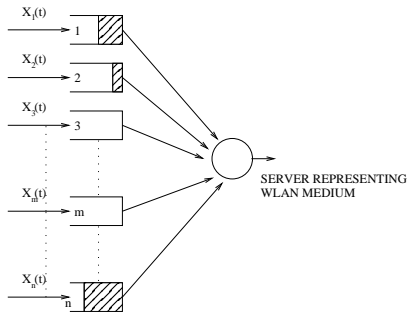}
\caption {System model}
\label{fig:model}
\end{figure}

We consider a single cell wireless local area network with $n$ nodes and no hidden nodes. We assume that the packet arrival process at node $i$ is Poisson with rate $\lambda_i$. Each node in the network shares the medium and uses the IEEE 802.11 DCF to exchange data with one another. A packet transmitted by a node is destined for any of the other nodes. Each node is assumed to have a single network output queue. All wireless links are assumed to operated at the same date rate.

From \cite{bianchi}, \cite{tay} and \cite{akumar}, it is known that the average aggregate rate of data transfer is dependent on the number of nodes contending. We also note that each node has equal probability of success. As in \cite{tobagi}, we model the network as a multiple queue single shared server system, where the service rate of the server is dependent on the number of non-empty queues and the server selects a non-empty queue uniformly at random. We assume that the destination can push data out of the network instantaneously. In this system, we are interested in quantifying the average delay between a packet's arrival to a queue and its departure from the system. 

\section{\large{Delay Modelling}}
\label{analysis}

\subsection{Computing aggregate service rate in 802.11 DCF}

We start this section by discussing the mechanism of 802.11 DCF. In 802.11 DCF, at the end of every successful transmission, there is \emph{short inter-frame space} (SIFS) which allows the receiving node to turn around its radio and send back a MAC level ACK packet. When the ACK transmission end, the channel is sensed to be idle by each of the node and each  one of them starts a DCF inter-frame space timer (DIFS). When the DIFS timer expires, each node enters a back-off phase. To prevent collisions, random back-off is used to order the transmissions. Each of the nodes freeze its back off timer in case any one of the nodes start transmission. Upon completion of transmission, the remainder of the back off is continued by the other nodes. If two or more nodes finish their back-offs within one slot of each other, a collision occurs. On detection of collision, the colliding  nodes sample back-offs from a doubled collision window. A successful transmission occurs if and only if exactly one of the nodes finishes its back-off in a given slot. Soon after, the node starts sending the packet. Upon sensing activity on the channel, other nodes freeze their back-off timers and defer their transmissions until the channel is sensed to be idle again.

In our model, atmost $n$ nodes contend for access to the wireless medium. Let $\beta_n$ be the probability that a node attempts transmission, when $n$ nodes are contending for access to the wireless medium. From \cite{bianchi}, $\beta_n$ can be expressed in terms of the conditional collision probability $p$ in two ways as follows.

\begin{eqnarray}
\beta_n(p) &=& \frac{2 \cdot (1-2p)}{(W+1) \cdot (1-2p) + pW \cdot (1 - {(2p)}^m)} \label{eq:beta1} \\
\beta_n(p) &=& 1 - (1-p)^{\frac{1}{n-1}} \label{eq:beta2}
\end{eqnarray}

Now, the RHS of Equation \eqref{eq:beta1} is monotonically decreasing from $\frac{2}{W+1}$ to $\frac{2}{2^mW+1}$, for $p \in [0,1]$, and the RHS of Equation \eqref{eq:beta2} is monotonically increasing from $0$ to $1$, for $p \in [0,1]$. We can use fixed point analysis to obtain $\beta$ for a given $n$. Similar approach has been followed in \cite{akumar}. Let $T_S, T_I$ and $T_C$ be the durations of success, idle and collision slots, respectively. Let us define the probability of a successful transmission ($p^{(n)}_S$), collision ($p^{(n)}_C$) and idle ($p^{(n)}_I$) as
$$ p^{(n)}_S = n\beta_n \cdot (1-\beta_n)^{n-1}$$
$$ p^{(n)}_I = (1-\beta_n)^n $$
$$ p^{(n)}_C = 1-  p^{(n)}_S - p^{(n)}_I$$	
By applying the Renewal Reward theorem, we define the average throughput in terms of packets per second as follows
$$S(n) = \frac{p^{(n)}_S}{p^{(n)}_IT_I + p^{(n)}_ST_S + p^{(n)}_CT_C}$$

\begin{figure}[h]
\centering
\includegraphics[scale=0.3,angle=-90]{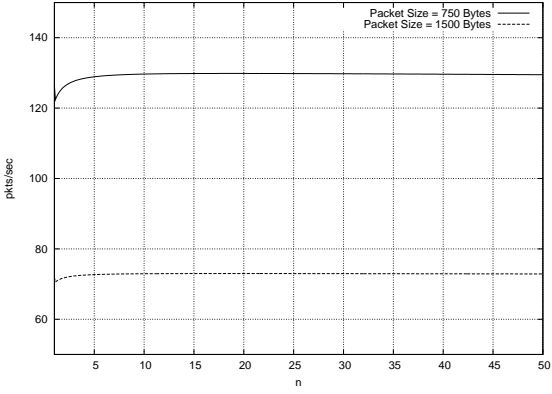}
\caption {Variation of throughput with the number of contending nodes. For the above plot, packets size was set as $1500\,Bytes$ and a transfer rate of $1\,Mbps$ and the parameters were set to mimic 802.11b}
\label{fig:thruput}
\end{figure}

From Figure \ref{fig:thruput}, it can be seen that the throughput remains fairly constant over a wide range of $n$. This value is taken as $C$ pkts/sec.

\subsection{Random polling system framework (RPS)}
We consider the system as $n$ infinite-buffer queues being served by a single server. When the server visits a non empty queue, it serves exactly one packet and moves onto the next queue. This type of polling system is classified as a \emph{1-Limited Radom Polling System}. The expected offered load to the system due to queue $i$ is defined as $\rho_i = \frac{\lambda_i}{C}$. The expected system utilization factor is define as $\rho = \frac{\sum^{n}_{i=1} \lambda_i}{C}$. Clearly, a necessary condition for the system to reach a stable state is 
$$\rho < 1$$ 

Now we restate the notations and assumptions used in \cite{lee}, merely for the purpose of understanding our simplifications. The polling policy is memoryless such that the next queue, say queue $j$, is selected for service with probability $\gamma_j$, where $0 < \gamma_j < 1$. The service time provided (if any) and the switchover time that follows are collectively defined as a \emph{period}. The service time of customers in queue $i$ is a nonnegative random variable with mean $p_i$ and second moment $p^{(2)}_i$. The switchover time after the server visits queue $i$ is a nonnegative random variable with mean $s_i$ and second moment $s^{(2)}_i$. Let $s = \sum^{n}_{j=1} s_j \gamma_j$. 

From \textit{Theorem 12} of \cite{lee}, we can obtain a closed form expression for the average waiting times for the packets at node $i$ as
\begin{eqnarray} \label{eq:ewi}
E[W_i] = \frac{E[Q_i]}{\lambda_i \cdot P\{Q_i \geq 1 \}} - \left(\frac{1-\rho_i}{\lambda_i}\right) 
\end{eqnarray}
with
\begin{eqnarray} \label{eq:eqi}
E[Q_i] = \frac{\psi_i}{\chi_i} + \frac{s  \lambda^2_i}{2\gamma_i (1-\rho) \chi_i} \cdot \frac{\sum^{n}_{l=1} \displaystyle \frac{p_l \psi_l}{\chi_l}}{1 - \sum^{n}_{l=1} \displaystyle \frac{p_l \lambda^2_l s}{2\gamma_l(1-\rho)\chi_l}}  
\end{eqnarray}
\noindent
where
\begin{eqnarray} 
\chi_i &=& 1 - \frac{s \lambda_i}{\gamma_i} - \frac{\rho s \lambda_i}{2\gamma_i(1-\rho)} \label{eq:chi} \nonumber \\
\psi_i &=& \frac{\nabla_{ii}}{2 \gamma_i} + \frac{\lambda_i}{2\gamma_i(1-\rho)} \cdot \sum^{n}_{l=1} p_l \nabla_{il} \label{eq:psi} \nonumber
\end{eqnarray}
Since our packet arrival processes are uncorrelated Poisson processes, we have a simplified definition of $\nabla_{ij}$ as
\begin{eqnarray} \label{eq:nabla}
\nabla_{ij} = \lambda_i \lambda_j \sum^{n}_{j=1} \left[ \gamma_j s^{(2)}_j + (p^{(2)}_j + 2s_jp_j)\frac{\lambda_j s}{(1-\rho)} \right] \\
+ \frac{2\lambda_is}{(1-\rho)}\boldsymbol 1_{\{ i=j\}} - \frac{\lambda_i\lambda_js (s_i + s_j + p_i + p_j)}{(1-\rho)} \nonumber
\end{eqnarray}
The probability of a queue being non-empty is given by
\begin{eqnarray} \label{eq:pqi}
P\{Q_i \geq 1 \} = \frac{s\lambda_i}{\gamma_i(1-\rho)} 
\end{eqnarray}
The above analysis applies only for a system with non-zero switch over time. 

\subsection{Application of RPS to Single Cell 802.11 WLAN}
According to our modelling assumptions, if the server visits a node with a nonempty queue, it will serve exactly one packet and will move to the next queue immediately (i.e, zero switchover time). Else, if the server visits a node with an empty queue, we assume that it will move to the next queue instantaneously (i.e, zero switchover time). We assume fair server allocation policy (i.e $\forall i\, , \gamma_i = \frac{1}{n}$). This emulates the process of the nonempty queue succeeding with equal probability. So we are interested in simplifying Equation \eqref{eq:ewi} to reflect the case of zero length switchover period. In a different context, the author in \cite{levy} has proposed that the expression for mean delay with zero switchover time can be obtained from the expression for non-zero switchover time by proper application of limits to the switchover time distribution. Motivated by this, we follow as similar approach to arrive at the expression for zero switchover time by defining the switchover time after servicing queue $i$ (i.e $s_i$) as a small constant value $\epsilon$. The expression for average delay when \emph{switch over times are zero} ($E[W^0_i]$) can be obtained by letting $\epsilon$ go to $0$. Thus we have 
\begin{eqnarray} \label{eq:ewi1}
E[W^0_i] =  \frac{\rho_i}{\lambda_i} - \frac{1}{\lambda_i} + \lim_{\epsilon \to 0} \frac{E[Q_i]}{\lambda_i \cdot P\{Q_i \geq 1 \}}
\end{eqnarray}
Substituting Equation \eqref{eq:pqi} in Equation \eqref{eq:ewi1}, we get
\begin{eqnarray} \label{eq:eq1}
E[W^0_i] =  \frac{\rho_i}{\lambda_i} - \frac{1}{\lambda_i} + \frac{\gamma_i(1-\rho)}{\lambda^2_i} \cdot \left( \lim_{\epsilon \to 0} \frac{E[Q_i]}{\epsilon} \right) \nonumber
\end{eqnarray}
The second term on the RHS in Equation \eqref{eq:eqi} becomes zero as $\epsilon \to 0$. Thus we have 
\begin{eqnarray} \label{eq:eq1}
\lim_{\epsilon \to 0} \frac{E[Q_i]}{\epsilon}  = \lim_{\epsilon \to 0} \frac{\frac{\psi_i}{\epsilon}}{ \chi_i}
\end{eqnarray}
It is easy to see that 
$$\lim_{\epsilon \to 0} \chi_i = 1 $$
Substituting Equation \eqref{eq:nabla} in Equation \eqref{eq:eq1}, we get
\begin{eqnarray} \label{eq:eq2}
\lim_{\epsilon \to 0} \frac{E[Q_i]}{\epsilon}  = \frac{1}{2\gamma_i} \left[ \lim_{\epsilon \to 0} \frac{\nabla_{ii}}{\epsilon} + \frac{\lambda_i}{(1-\rho)} \sum^{n}_{l=1} p_l \cdot \lim_{\epsilon \to 0} \frac{  \nabla_{il}}{\epsilon} \right]
\end{eqnarray}
From the arguments in the previous section, we assume  that a packet is serviced at a constant time of $\frac{1}{C} $ sec. By adopting the throughput computation by Bianachi \cite{bianchi}, we have abstracted the idle and the collision slots into the constant service time for a packet given by $\frac{1}{C}$ sec. We model our system using the RPS framework by taking a server which serves at fixed rate. i.e
$$\forall i,\, p_i = \frac{1}{C} \textrm{ and } p^{(2)}_i = \frac{1}{C^2}$$ 
After substituting the values for $p_i$ and $p^{(2)}_i$ in Equation \eqref{eq:nabla} and by taking the limit $\epsilon \to 0$, we get
\begin{eqnarray} \label{eq:nablaii}
\lim_{\epsilon \to 0} \frac{\nabla_{ii}}{\epsilon} = \frac{2\lambda_i}{(1-\rho)} + \frac{\lambda^2_i(\rho-2)}{C(1-\rho)} 
\end{eqnarray}
Similarly, for $i \neq j$ we get
\begin{eqnarray} \label{eq:nablaij}
\lim_{\epsilon \to 0} \frac{\nabla_{ij}}{\epsilon} = \frac{\lambda_i\lambda_j\rho}{C(1-\rho)} - \frac{2\lambda_i\lambda_j}{C(1-\rho)} 
\end{eqnarray}
Now, setting $p_l=\frac{1}{C}$ and using Equations \eqref{eq:nablaii} and \eqref{eq:nablaij}, we obtain
\begin{eqnarray} \label{eq:eq3}
\sum^{n}_{l=1} p_l \cdot \lim_{\epsilon \to 0} \frac{\nabla_{ij}}{\epsilon} = \frac{\lambda_i(\rho^2-2\rho+2)}{C(1-\rho)} 
\end{eqnarray}
Substituting Equations \eqref{eq:eq3} and \eqref{eq:nablaii} in Equation \eqref{eq:eq2}, we get
\begin{eqnarray} \label{eq:eqnl}
\lim_{\epsilon \to 0} \frac{E[Q_i]}{\epsilon} =  \frac{1}{2\gamma_i} \left(\frac{2\lambda_i}{(1-\rho)} + \frac{\lambda^2_i\rho}{C(1-\rho)^2}\right)
\end{eqnarray}
Substituting Equation \eqref{eq:eqnl} in Equation \eqref{eq:eq1}, we obtain the closed form expression for average delay at node $i$ as
\begin{eqnarray} \label{eq:delay}
E[W^0_i] =  \frac{(2-\rho)}{2C(1-\rho)}
\end{eqnarray}

\subsection{Discussion}
We observe that the expression for average delay in queue $i$ (i.e, Equation \eqref{eq:delay}) is the same for all the queues, irrespective of the nonhomogeneous arrival process. The expression for the average delay depends only on aggregate throughput of IEEE 802.11 DCF ($C$) and on the expected system utilization factor ($\rho$).

The invariance in the mean delay across the queues, even when the packet arrival process is nonhomogeneous Poisson process, can be explained by the \emph{``Poisson Arrivals See Time Averages (PASTA)''} property of the packet arrival process. 

\section{\large{Validation}}
\label{sim}
In this section, we show the plots and tables which compare the analytical against the simulated values of mean delay. The simulation was done using Qualnet 4.5, which is a discrete event simulation system. In order to obtain accurate estimates of the mean delay, the simulation was run long enough so that the average delay at the root was within $1 \mu s$ interval. This process was continued for $30$ simulation runs to obtain a confidence interval for the mean delay with $95\%$ confidence. For the analytical computation of mean delay, the value of $C$ is taken as 72.5 packets/sec.

\begin{table}[!h] \label{table1}
\caption{Table of delay for simulation and the analytical values for aggregate rate $\approx \, 60 \, packets/sec$. Packet size is $1500 \, Bytes$. Data Rate is $1 \, Mbps$. $3$ contending nodes}
\begin{center}
\begin{tabular}{|c|c|c|c|c|c|c|}
\hline $\lambda_1 $ & $\lambda_2$ & $\lambda_3$ & $d^{(1)}_{avg}$ & $d^{(2)}_{avg}$ & $d^{(3)}_{avg}$ & $E[W^0]$  \\ 
\hline 10.0 & 30.3 & 20.0 & 42.4 & 42.6 & 42.1 & 47.9 \\
\hline 20.0 & 20.0 & 20.0 & 43.9 & 45.5 & 43.5 & 46.9 \\
\hline 2.0 & 29.4 & 29.4 & 42.6 & 40.5 & 43.5 & 49.7 \\
\hline 1.0 & 1.0 & 58.8 & 34.6 & 35.9 & 34.9 & 49.7 \\
\hline
\end{tabular} 
\end{center}
\end{table}

\begin{table}[!h] \label{table2}
\caption{Table of delay for simulation and the analytical values for aggregate rate $\approx \, 30 \, packets/sec$. Packet size is $1500 \, Bytes$. Data Rate is $1 \, Mbps$. $3$ contending nodes}
\begin{center}
\begin{tabular}{|c|c|c|c|c|c|c|}
\hline $\lambda_1 $ & $\lambda_2$ & $\lambda_3$ & $d^{(1)}_{avg}$ & $d^{(2)}_{avg}$ & $d^{(3)}_{avg}$ & $E[W^0]$  \\ 
\hline 10.0 & 10.0 & 10.0 & 18.8 & 18.4 & 18.3 & 18.7 \\
\hline 5.0 & 10.0 & 14.9 & 18.5 & 18.4 & 18.3 & 18.6 \\
\hline 1.0 & 1.0 & 27.8 & 18.9 & 18.5 & 18.3 & 18.6 \\
\hline 5.0 & 12.5 & 12.5 & 18.5 & 18.3 & 18.2 & 18.6 \\
\hline 1.0 & 1.0 & 58.8 & 34.6 & 35.9 & 34.9 & 49.7 \\
\hline
\end{tabular} 
\end{center}
\end{table}

\begin{table}[!h] \label{table3}
\caption{Table of delay for simulation and the analytical values for aggregate rate $\approx \, 60 \, packets/sec$. Packet size is $1500 \, Bytes$. Data Rate is $1 \, Mbps$. $4$ contending nodes}
\begin{center}
\begin{tabular}{|c|c|c|c|c|c|c|c|c|}
\hline $\lambda_1 $ & $\lambda_2$ & $\lambda_3$ & $\lambda_4$ & $d^{(1)}_{avg}$ & $d^{(2)}_{avg}$ & $d^{(3)}_{avg}$ & $d^{(4)}_{avg}$ & $E[W^0]$  \\ 
\hline 14.9 & 14.9 & 14.9 & 14.9 & 44.1 & 43.9 & 44.1 & 43.3 & 45.9 \\
\hline 1.0 & 19.6 & 19.6 & 19.6 & 41.7 & 42.7 & 41.5 & 42.7 & 46.3 \\
\hline 7.5 & 12.5 & 17.5 & 22.2 & 42.1 & 42.0 & 41.3 & 41.2 & 46.2 \\
\hline 1.5 & 1.5 & 1.5 & 55.5 & 30.5 & 32.7 & 32.2 & 32.4 & 46.8 \\
\hline
\end{tabular} 
\end{center}
\end{table}

\begin{table}[!h] \label{table4}
\caption{Table of delay for simulation and the analytical values for aggregate rate $\approx \, 30 \, packets/sec$. Packet size is $1500 \, Bytes$. Data Rate is $1 \, Mbps$. $4$ contending nodes}
\begin{center}
\begin{tabular}{|c|c|c|c|c|c|c|c|c|}
\hline $\lambda_1 $ & $\lambda_2$ & $\lambda_3$ & $\lambda_4$ & $d^{(1)}_{avg}$ & $d^{(2)}_{avg}$ & $d^{(3)}_{avg}$ & $d^{(4)}_{avg}$ & $E[W^0]$  \\ 
\hline 7.5 & 7.5 & 7.5 & 7.5 & 18.9 & 18.7 & 18.5 & 18.3 & 18.6 \\
\hline 3.7 & 6.3 & 8.7 & 11.1 & 18.7 & 18.5 & 18.5 & 18.4 & 18.6 \\
\hline 0.5 & 9.8 & 9.8 & 9.8 & 19.1 & 18.8 & 19.3 & 18.9 & 18.6 \\
\hline 0.5 & 0.5 & 0.5 & 27.8 & 20.7 & 20.3 & 20.6 & 20.4 & 18.5 \\
\hline
\end{tabular} 
\end{center}
\end{table}

In Tables \ref{table1},  \ref{table2}, \ref{table3} and \ref{table4}, we validate the invariance of mean delay across queues under nonhomogeneous Poisson arrivals. The delay values in the table are in units of $milliseconds$. Tables \ref{table1} and  \ref{table3}, illustrate the variation of delay when the aggregate arrival rate is held constant at $60 \, packets/sec$, while Tables \ref{table2} and  \ref{table4} show the variation for an aggregate arrival rate of $30 \, packets/sec$. It can be observed that for light to moderately loaded systems, the analytical and simulated values are in excellent agreement with each another. 

\begin{figure}[!h]
\centering
\includegraphics[scale=0.3,angle=-90]{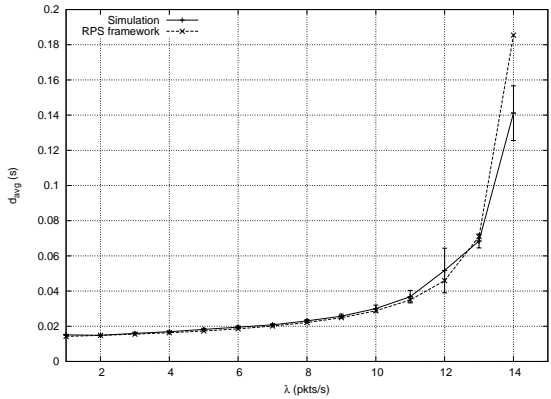}
\caption{Variation of $d_{avg}$ with $\lambda$ when 5 nodes are sending data to some destination. Packet size is $1500\,Bytes$ and the transmission rate is $1\,Mbps$ }
\label{fig:plot5}
\end{figure}

In Figure \ref{fig:plot5}, we plot the variation in $d_{avg}$ against arrival rate $\lambda$. We observe that for entire capacity region (i.e, $\lambda < 14.56$ ), the theoretical mean delay and simulated mean delay are close to each other 

\begin{figure}[!h]
\centering
\includegraphics[scale=0.3,angle=-90]{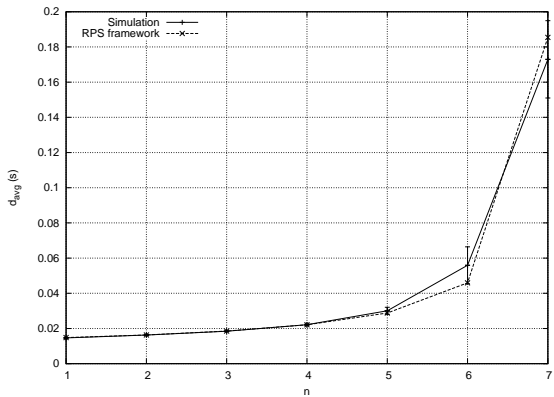}
\caption {Variation of $d_{avg}$ with $n$ when $\lambda =10$ pkts/sec. Packet size is $1500\,Bytes$ and the transmission rate is $1\,Mbps$ }
\label{fig:rate5}
\end{figure}

In Figure \ref{fig:rate5} , we plot the variation in $d_{avg}$ against the number of contending nodes $n$ for a given arrival rate. In Figure \ref{fig:rate5} packets of size $1500\, Bytes$ were used. We observe that in both the Figures, the theoretical and simulated means are very close to each other. We can also observe that with larger sized packets, the Random Polling System (RPS) framework is able to model single hop wireless mesh network with reasonable accuracy.

\section{\large{Conclusion and Future Work}}
In this paper, we have proposed a Random Polling System(RPS) framework for mean delay for single-hop wireless mesh networks. We have obtained a simple closed form expression for mean delay for the case of Poisson packet arrival. 

We have abstracted the mechanism of IEEE 802.11 DCF by using results from \cite{bianchi}. This abstraction has enabled us to model the delay in a single cell IEEE 802.11 wireless area network. Our analysis enable us to compute the mean delay for homogeneous Poisson arrivals from very simple expression as compared to the iterative approach in \cite{panda}. It can be seen that, even for the scenario of nonhomogeneous packet arrivals, the mean delay across the queues remain same and moreover depends only on the aggregate arrival rate and the saturated throughout of the IEEE 802.11 MAC. Simulations indicate that the proposed framework is able to model the mean delay in single hop wireless mesh network with reasonable accuracy, provided the system operates within the capacity region. 

Our ongoing work is concerned with refining the model; we are seeking a tractable and simple approach with which we can model user delay in single hop wireless mesh network with even more accuracy. Subsequently, we would like to extend our framework to arrive at either closed form expressions or numerically solvable equations for non-Poisson arrivals. We would also like to extend this to multi-hop wireless mesh networks by studying the behavior of hierarchical polling systems.

\bibliography{delay_modelling_rps}

\end{document}